\documentclass{article}%
\usepackage{amsmath}
\usepackage{amsfonts}
\usepackage{amssymb}
\usepackage{graphicx}%
\begin{document}

\title{Soliton Formation in Neutral Ion Gases: Exact Analysis}
\author{Babur M. Mirza\thanks{E-mail address: bmmirza2002@yahoo.com}\\Department of Mathematics, \\Quaid-i-Azam University, \\Islamabad. 45320. \\Pakistan.}
\maketitle

\begin{abstract}
It is shown here that in neutral ion gases the thermal energy transport can
occur in the form of new types of thermal soliton waves. The solitons can form
under a vanishing net heating function, and for a quadratic net heating. It is
predicted that these solitons play an important role in \ a diversity of
terrestrial and astrophysical phenomena. We claim that the reported soliton
waves can be observed under ordinary laboratory conditions.

\end{abstract}


\section{Introduction}

Compressive wave-front propagation is one of the major causes of energy
transport in thermally conducting gases, and gives rise to a diverse range of
physical phenomena (see [1-3] for further references). It is known that except
for the case of subsonic wave motion, all types of associated temperature
perturbation cause thermal instability in such systems. This behavior is known
as the Field criterion [4], which allows only highly subsonic wave motion in a
thermally stable medium. A direct consequence of the Field criterion is that
the pressure variations remain small, hence can be neglected in the analysis
of compressive thermal waves. The resulting waves formation is highly
nonlinear, and depends on the heating function. Linearization and phase space
analysis have been used to determine the nature of such waves, which have
indicated that apart from travelling wave formations, there are steady wave
fronts in such cases.

In this work we show that in thermally conducting ion gases, the compressive
wave propagation can occur in the form of stable, shape preserving wave forms,
known as soliton waves. Solitons can be identified as travelling wave
solutions that retain their amplitude throughout their propagation, and have
constant amplitude in the asymptotic limits. We show that these soliton waves
are very general feature of thermally conducting gases, both in astrophysical
and under normal laboratory conditions. These play an important role in the
energy transport in the media. We state the conditions under which such
soliton waves can form in a neutral ion gas, under a linear or a quadratic or
a zero net heating function.

\section{Basic Equations}

The propagation of a wave in an initially uniformly distributed gas, is
governed by the basic equations of mass, momentum, and thermal energy
conservation conservation. With gas density $\rho$, at pressure $p$, and
temperature $T$ these equation are:%
\begin{equation}
\frac{\partial\rho}{\partial t}+\mathbf{\nabla}\cdot(\rho\mathbf{v})=0,
\label{1}%
\end{equation}%
\begin{equation}
\rho\frac{D\mathbf{v}}{Dt}+\nabla p=0, \label{2}%
\end{equation}%
\begin{equation}
\left(  \frac{1}{1-\gamma}\right)  \frac{Dp}{Dt}-\left(  \frac{\gamma
}{1-\gamma}\right)  \frac{p}{\rho}\frac{D\rho}{Dt}=\mathbf{\nabla}%
\cdot(\lambda\mathbf{\nabla}T)+Q(\rho,T), \label{3}%
\end{equation}
where $\mathbf{v}$ is the velocity of the wave in the gas, and $Q(\rho,T)$ is
the net heat gain per unit volume per unit time. It expresses the difference
between the heating and cooling rates of the medium. Here $\lambda$ is the
thermal conductivity coefficient, given by the ratio of the of specific heats
$\gamma=c_{p}/c_{v}$.

For a gas at fixed volume and mass, the net heat gain is a function of
temperature only. It therefore follows from equation (1) to (3) that
instability occurs for the gaseous medium when $\left(  \frac{\partial
Q}{\partial\ln T}\right)  _{p}>0$ (Field 1965). This condition is the Field
stability criterion. In one spatial dimension the Field criterion implies that
wave motion of heating or cooling interfaces must be highly subsonic. This
means that we can neglect the velocity term in the momentum equation, hence
obtain from equation (2):%
\begin{equation}
\frac{\partial p}{\partial x}=0. \label{4}%
\end{equation}
implying that pressure is constant within the medium. A change to Lagrange
variable $\eta(x,t)$ defined by $\eta(x,t)=\int_{-\infty}^{x}\rho(x^{\prime
},t)dx^{\prime}$, gives for the mass conservation equation (1):%
\begin{equation}
\frac{\partial\eta}{\partial t}\mid_{x}=-\rho v, \label{5}%
\end{equation}
where $v$ is the velocity in $x$-direction. This gives $\partial/\partial
x\mid_{t}=\rho\partial/\partial\eta\mid_{t}$and $\partial/\partial t\mid
_{x}=\partial/\partial t\mid_{\eta}-\rho u\partial/\partial\eta\mid_{t}$. The
Lagrange derivative $D/Dt$ can now be expressed as:%
\begin{equation}
\frac{\partial f}{\partial t}\mid_{x}+v\frac{\partial f}{\partial x}\mid
_{t}=\frac{\partial f}{\partial t}\mid_{\eta}. \label{6}%
\end{equation}
Using these results in the energy equation (3) we obtain,%
\begin{equation}
\frac{1}{\gamma-1}\frac{Dp}{Dt}-\frac{\gamma}{\gamma-1}\frac{p}{\rho}%
\frac{D\rho}{Dt}=\frac{\partial}{\partial x}\left(  \lambda\frac{\partial
T}{\partial x}\right)  +Q, \label{7}%
\end{equation}
which becomes%
\begin{equation}
-\frac{\gamma}{\gamma-1}\frac{p}{\rho}\frac{\partial\rho}{\partial t}%
\mid_{\eta}=\rho\frac{\partial}{\partial\eta}\left(  \lambda\rho\frac{\partial
T}{\partial\eta}\right)  +Q. \label{8}%
\end{equation}
In the subsonic case the equation of state is $p=\left(  \mathcal{R}%
/\mu\right)  \rho T$, it therefore follows that $\rho\ $is proportional to
$1/T$. Thus both $\lambda$ and $Q$ are functions of temperature only. Let now
the new time variable $\tau=\left(  (\gamma-1)p/\gamma\left(  \mathcal{R}%
/\mu\right)  ^{2}\right)  t$, and denote $\mathcal{L(}T)=Q(T)\left(
\frac{\mathcal{R}/\mu}{p}\right)  ^{2}/T$. Then equation (10) gives the
governing equation for the evolution of the temperature front in a thermally
conducting gas:%
\begin{equation}
\frac{\partial T}{\partial\tau}=\frac{\partial}{\partial\eta}\left(
\frac{\lambda(T)}{T}\frac{\partial T}{\partial\eta}\right)  +\mathcal{L(}T).
\label{9}%
\end{equation}

\section{Soliton Wave Analysis}

The conductivity function $\lambda(T)$ for a neutral ion gas is proportional
to$\sqrt{T}$. In the case of steadily moving wave front in $\eta$ space, with
velocity $V$, we set $\xi=c(\eta-K\tau)$. This gives in equation (12):%
\begin{equation}
c^{2}u^{-1}\frac{d^{2}u}{d\xi^{2}}+cK\frac{du}{d\xi}+F(u)=0, \label{10}%
\end{equation}
where $u=\sqrt{T}$, and $F(u)=\frac{1}{2}T^{-1/2}\mathcal{L(}T)$ is the net
heating function. We now consider equation (13) for the case of linear, and
quadratic net heating functions.

If the net heating function is linear, we have from equation (10) the
nonlinear wave equation:%
\begin{equation}
c^{2}\frac{1}{u}\frac{d^{2}u}{d\xi^{2}}+cK\frac{du}{d\xi}+Au+B=0. \label{11}%
\end{equation}
We seek the \ solution to this equation \ such that $u^{\prime}(\xi)$ is a
function of the separable form,%
\begin{equation}
u^{\prime}(\xi)=a^{2}-u(\xi)^{2}. \label{12}%
\end{equation}
This implies that%
\begin{equation}
u^{\prime\prime}(\xi)=-2a^{2}u(\xi)+2u(\xi)^{3}. \label{13}%
\end{equation}
Therefore equation (11) becomes after substitution from equations (12) and
(13), and collecting terms of the like powers in $u$:%
\begin{equation}
(2c^{2}-cK)u^{3}+Au^{2}+(cKa^{2}+B-2a^{2}c^{2})u=0. \label{14}%
\end{equation}
Equating the coefficients of $u^{2}$, $u^{1}$, and $u^{0}$ we get%

\begin{equation}
K=2c,\text{ }A=0\text{, }B=0. \label{15}%
\end{equation}
This case correspond to the vanishing net heating function, and from equations
(12) and (15) the solution can be written as%
\begin{equation}
u(\xi)=a\tanh a(\xi+\xi_{0}), \label{16}%
\end{equation}
where $\xi_{0}$ corresponds to the initial value of the travelling wave
variable. Equation (16) gives for the temperature
\begin{equation}
T(\eta,\tau)=a^{2}\tanh^{2}ac(\eta-2c\tau+\xi_{0}/c). \label{17}%
\end{equation}
Equation (17) implies that as $\xi\rightarrow\pm\infty$, $T\rightarrow
T_{\pm\infty}=a^{2}$. The aysmptotic limit of temperature must correspond to
the equilibrium state. We re-scale the temperature as $\tilde{T}(\eta,\tau)$
$=T_{\pm\infty}-T(\eta,\tau)$, so that equation (17) becomes%

\begin{equation}
\tilde{T}(\eta,\tau)=T_{\pm\infty}\left(  \operatorname*{sech}ac(\eta
-2c\tau+\xi_{0}/c)\right)  ^{2}. \tag{17a}\label{17a}%
\end{equation}
This represents a free soliton wave solution when $F(u)=0$. Physically this
corresponds to the case when heating of the gas occurs at the same rate as
cooling. The soliton amplitude in this case is $a^{2}$ and its speed is
$2ac^{2}$. As $\xi\rightarrow\pm\infty$, the soliton attains a fixed amplitude
$a^{2}$, whereas it retains its profile during propagation. Figure (1) shows
plot of the soliton (17a) as function of variables $\eta$ and $\tau$.

For a quadratic heating function $F(u)=Au^{2}+Bu+C$, we have by a similar
procedure:%
\begin{equation}
(A+2c^{2}-cK)u^{3}+Bu^{2}+(cKa^{2}+C-2a^{2}c^{2})u=0. \label{18}%
\end{equation}
Equating coefficients of the various powers of $u$, gives the determining
equations,%
\begin{equation}
A+2c^{2}-cK=0,\text{ }cKa^{2}+C-2a^{2}c^{2},\text{ }B=0, \label{19}%
\end{equation}
we get%
\begin{equation}
c=\frac{K}{4}\pm\frac{1}{4}\sqrt{K^{2}-8A},\text{ }a=\pm\sqrt{C/A},\text{
}B=0, \label{20}%
\end{equation}
Thus corresponding to the net heating function $Au^{2}+C$ we have the soliton
solution%
\begin{equation}
u(\xi)=\pm\sqrt{C/A}\tanh\pm\sqrt{C/A}[c(\eta-K\tau)+\xi_{0}], \label{21}%
\end{equation}
or in terms of the temperature variable%
\begin{equation}
T(\eta,\tau)=\frac{C}{A}\tanh^{2}\pm\sqrt{\frac{C}{16A}}[(K+\sqrt{K^{2}%
-8A})(\eta-K\tau)+4\xi_{0}]. \label{22}%
\end{equation}
Re-scaling the temperature as $\tilde{T}(\eta,\tau)$ $=T_{\pm\infty}%
-T(\eta,\tau)$, where $T_{\pm\infty}=C/A$, we obtain%

\begin{equation}
T(\eta,\tau)=T_{\pm\infty}\operatorname*{sech}^{2}\pm\sqrt{\frac{C}{16A}%
}[(K+\sqrt{K^{2}-8A})(\eta-K\tau)+4\xi_{0}]. \tag{22a}\label{22a}%
\end{equation}
Equation (22a) shows that the soliton solution for a purely quadratic function
has a strong amplitude dependence on the heating function parameters. In this
way it represents a totally distinct soliton solution than that for the
previous case. The thermal soliton (22a) has amplitude $C/A$, which increases
as the parameter $A$ decreases. Also the wave speed is proportional to
$\sqrt{C/16A}$, which shows that the soliton wave moves faster if the
quadratic term in the net heating function is adjusted to smaller values as
compare to the constant terms in the heating function. Another interesting
feature of the soliton solution (22a) is that the argument of the
$\operatorname*{sech}$ function involves the square root $\sqrt{K^{2}-8A}$,
which can become imaginary when $K^{2}<8A$, thus exhibit drastically different
behavior depending on the threshold value $K=\pm\sqrt{8A}$.

\section{Conclusions}

In this communication it is reported that thermally conducting gases in
neutral ionized states contain thermal solitons that participate in the
general process of energy transfer in the medium. For a given linear or
quadratic net heating function these soliton waves can form as single solitary
waves or, depending on the parameters of the heating function, can appear as
multiple soliton waves. In the quadratic case, the net heating function
parameters not only determine their amplitude but their speed in the gas as
well. The existence of free solitons shows that there can be energy transfer
via solitons even when the cooling of the gas occurs at the same rate as its
heating. In this case the amplitude as well as the speed of the soliton wave
depends on the equilibrium temperature of the gas. Since temperature is
inversely propotional to the density of the gas, the soliton effects can be
observed in density profile of a neutral ion gas, under appropriate conditions.

Neutral ion gases are most prevalent in astrophysical and atmospheric
conditions, such as the solar corona, interstellar medium, and galactic
clusters, ionosphere, upper atmosphere of the Earth and planets.

However the thermal solitons can also be observed under ordinary laboratory
conditions. It is predicted here that for the vanishing and quadratic net
heating functions, cases which are easier to be produced in a lab with
adjustable heating parameters for neutral ion gases, the existence of these
thermal solitons can be demonstrated. As far as the author knows this
experimental proof yet remains to be given.

\bigskip

Figure Caption:

Figure 1: A free soliton wave based on equation (17a), formed at equilibrium
temperature $T_{\pm\infty}=0.1$, $\xi_{0}=0$, and $c=0.2$.

\end{document}